\documentclass[aps,prl,twocolumn,showpacs,floatfix,superscriptaddress]{revtex4}
\usepackage{graphicx}
\usepackage{graphics}
\usepackage{epsfig}
\usepackage{amsmath}
\usepackage{amssymb}
\usepackage{dcolumn}
\usepackage{bm}
\usepackage{longtable}

\begin{document}

\title{Cold Fermionic Atoms in Two-Dimensional Traps -- 
Pairing versus Hund's Rule}

\author{M. Rontani}

\affiliation{CNR-INFM National Research Center S3, Via Campi 213/A, 41100
  Modena, Italy}

\email{rontani@unimore.it}
\homepage{www.nanoscience.unimore.it/max.html}

\author{J. R. Armstrong}

\author{Y. Yu}

\author{S.~{\AA}berg}

\author{S. M. Reimann}

\affiliation{Mathematical Physics, Lund Institute of
Technology, P.O. Box 118, 22100 Lund, Sweden}

\date{\today }

\begin{abstract}
The microscopic properties of few
interacting cold fermionic atoms confined in a 
two-dimensional (2D) harmonic trap are
studied by numerical diagonalization. For repulsive interactions, a
strong shell structure dominates, with Hund's rule acting at its
extreme for the mid-shell configurations. In the attractive case,
odd-even oscillations due to pairing occur 
simultaneously with 
deformations in the internal structure 
of the ground states, as seen from pair correlation
functions.
\end{abstract}


\maketitle Quantum dots and related nanostructures appeared as the
first ``artificial atoms''~\cite{artifiatoms}, where semiconductor
techniques allowed to design the quantum confinement of a
few particles in the laboratory (see~\cite{reimannRMP} for a
review). In more recent years, artificial quantum confinement also
became possible with ultracold trapped {\it atomic} gases 
\cite{books}: Magneto-optical techniques allow to
create very clean systems where the trapped particles can be bosons
as well as fermions. Strength and sign of their
interactions can be tuned, making them repulsive or attractive via
Feshbach resonances, allowing for a variety of quantum many-body
phenomena to be studied theoretically as well as experimentally.
Cold atoms may be loaded into ``deep'' optical lattices, where the 
single sites may resemble harmonic confinement.
With cold, trapped fermions, it became possible to experimentally
study the crossover from a weakly attractive Fermi gas to the unitary
limit of infinite scattering length,
and beyond~\cite{coldferm}. Attractive fermionic atoms in traps
share many of their properties with nuclei, such as
Bardeen-Cooper-Schrieffer (BCS) pairing and the
occurence of shell structure~\cite{mottelson}. 

The ``Coulomb blockade'' in the discrete charging
and discharging of quantum dots leads to significant
conductance oscillations. These are manifest in the ``fundamental
energy gap'' $\Delta _2 (N)= \Delta_1(N+1)-\Delta_1(N)$,
given by the difference between the modulus of the particle-removal and
the addition energy, i.e., by the difference in
the chemical potential of a quantum system confining $N+1$ and $N$ particles,
$\Delta_1(N)= E_0(N) -E_0(N-1)$, where
$E_0(N)$ is the $N$-body ground state energy.
Very recently, Cheinet {\it et al.}~\cite{cheinet2008}
reported on the realization of a similar interaction
blockade~\cite{capelle2007}  in the
transport of cold atoms through a double well in an optical lattice,
observing discrete steps in the well population with increasing ``bias''
potentials [note that the atom number distribution as a function 
of the bias relates to $\Delta _2(N)$].
These findings indeed bring implementations of ``atomtronics'',
as suggested in the pioneering work of Seaman {\it et al.}~\cite{seaman2007},
much closer to realization.

In this Letter, we thus
investigate effects of interaction blockade for quantum dot-like,
2D fermionic atom traps.
For repulsive interactions, a strong shell structure
occurs and Hund's rule acts at its extreme at half-filled shells.
For attractive interactions, however, a pronounced odd-even staggering
in the ground state energies (as recently discussed for 3D systems by Zinner
{\it et al.}~\cite{zinner2008})
is accompanied by broken symmetries in the
internal structure of the many-body states.


Let us now consider $N$ atoms of spin 1/2 and equal masses $m$ confined in a
2D harmonic trap, interacting through a contact potential,
\begin{equation}
H = \sum_{i=1}^{N}\left(-\frac{\hbar^2}{2m}\nabla_i^2+\frac{1}{2}m\omega_0^2
r_i^2\right)+\frac{1}{2}g^{\prime}\sum_{i\neq j}
\delta ^{(2)}(\bm{r}_i-\bm{r}_j),
\label{eq:H}
\end{equation}
where $g^{\prime}$ is the coupling constant. We use as energy unit
$\hbar\omega_0$ and as length unit 
$\ell = (\hbar/m\omega_0)^{1/2}$. The
dimensionless coupling constant is $g =
g^{\prime}/(\hbar\omega_0\ell^2)$. The few-body problem of
Eq.~(\ref{eq:H}) is solved by means of the full configuration
interaction (CI) method \cite{RontaniCI}. In more than one spatial
dimension, contact interactions do not allow to diagonalize the
many-body Hamiltonian unless regularized properly. An approximate
way to achieve this is a momentum-cutoff to a given
subspace~\cite{momentumcutoff},
effectively renormalizing the interaction
strength~\cite{rontani2008}. In the applied model space of the six
lowest oscillator shells we can then relate the interaction strength
$g$ to the 2D scattering length $a$ by comparing the CI
two-body ground-state energy with the exact result derived by
Busch and coworkers \cite{Busch}, as shown in Ref.~\onlinecite{rontani2008}. 
The obtained 
values of $a$ are listed in note \onlinecite{comment2}. 

Expanding the interacting $N$-body state on the given set of Slater
determinants and diagonalizing~\cite{Don}, the CI method provides
the many-body energies and wave functions of both ground states and
(low-lying) excitations, which are eigenstates of the total spin $S$
and $z$-projection of the orbital angular momentum $M$.
\begin{figure}
\centerline{\epsfig{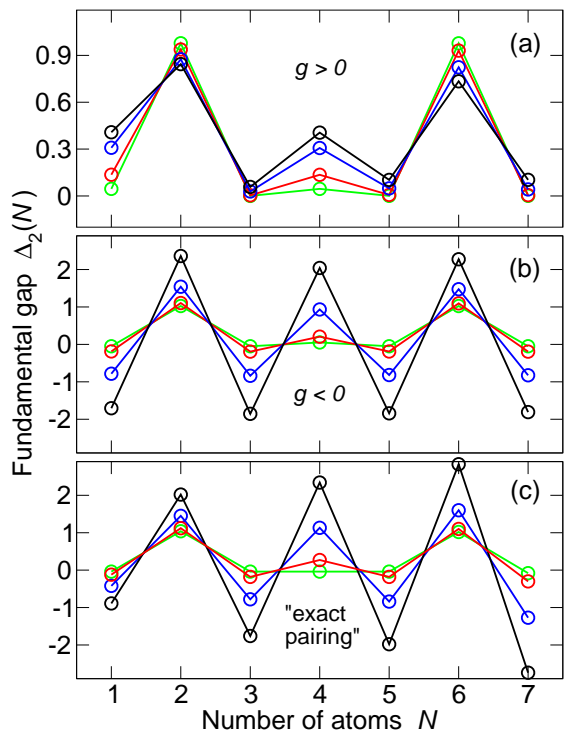}}
\caption{(Colour)  $\Delta_2(N)$ vs 
$N$. (a) The green, red, blue,
black curves are calculated by using $g=$ 0.3, 1, 3, 5.
(b) The green, red, blue, black curves are
calculated by using $g=$ -0.3, -1, -3, -5. 
(c) Results from the seniority model.
\label{addition}}
\end{figure}
Figure~\ref{addition} shows the fundamental gap $\Delta_2(N)$ vs
$N$. In the upper panel, for {\it repulsive} interactions $g>0$,
one clearly recognizes the distinctive features of a shell structure,
strikingly similar to those observed for quantum dots \cite{tarucha1996} 
(confirming the earlier prediction by Capelle {\it et al.}~\cite{capelle2007}).
The two peaks for $N=2,6$ form when the first two energy
shells of the 2D harmonic oscillator are completely filled
by atoms as they are consecutively loaded into the trap.
Since the height of the peaks
is mainly dictated by the energy spacing ($\hbar \omega _0=1$)
between neighboring shells, $\Delta_2(N=2,6)\approx 1$.
$\Delta _2(N)$
depends only weakly on the interaction strength $g$,
as it increases from $g=0.3$ (green curve) up to $g=5$ (black curve).
The peak for $N=4$ at mid-shell between $N=2$ and $N=6$
is a consequence of Hund's rule acting at its extreme,
where up to half-filling of the degenerate shell
the  Pauli principle eliminates the interactions.

The signatures of {\it attractive} interaction ($g<0$) are
dramatically different from the repulsive case of
Fig.~\ref{addition}(a). Figure \ref{addition}(b) shows that the
non-interacting shell structure is gradually washed out as the
magnitude of atom-atom attractive interaction is increased. As $g$
goes from $g=-0.3$ (green curve) up to $g=-5$ (black curve),
$\Delta_2(N)$ develops a pronounced even-odd pattern, with
$\Delta_2>0$ ($\Delta_2<0$) for even (odd) $N$, corresponding to
$S=0$ ($S=1/2$). Pairing effects are likely to be the dominant
mechanism behind these odd-even oscillations. If $N$ is even, all atoms
form singlet pairs and a large amount of energy is required to add
one unpaired atom. For odd $N$, $\Delta_2(N)<0$ since
energy is gained by pairing with an opposite-spin particle.
However, the odd-even staggering of Fig.~\ref{addition}(b)
could originate from either pairing or Jahn-Teller effects, 
as discussed for 
nuclei, metallic grains, clusters~\cite{Jahn-Teller},
and even for 2D systems~\cite{reimann1998}.

In order to isolate effects from pairing we study the seniority
model from nuclear physics (also called the method of 'exact
pairing'~\cite{volya2001}), applied to the 2D system. The model
restricts the interactions in the 2D trap to be exclusively of the
pairing type, that act only between particles in time-reversed
orbits. Unpaired particles do not participate in the interaction
beyond the mean-field level, except for blocking certain final
states available for pairing. Using an attractive $\delta$-function
pairing potential as for the full microscopic model above, we truncate
the model space after the fourth oscillator shell. The results are
shown in Fig.~\ref{addition}(c), comparing well with the full
many-body calculations. For weak interactions, $g=-0.3$, 
$\Delta_2(N)$ displays the non-interacting shell structure. 
For stronger $g$, a pronounced odd-even staggering
appears. We find large gaps between the ground and first excited
state in even--, and the absence of a gap in odd-numbered systems.
To discern between inter- and intra-shell pairing \cite{mottelson},
we study the fraction of the pairing energy that comes from the
off-diagonal Hamiltonian matrix elements, which involve pairs being
moved between the oscillator shells. For smaller interaction strengths, this
energy is a fraction of a percent. Increasing the coupling to
$g=-5$, the off-diagonal contribution is larger than 20\%. This
was roughly the same for all particle numbers, though partially
filled shells have higher contributions than particle numbers around
closed shells.
\begin{figure}
\centerline{\epsfig{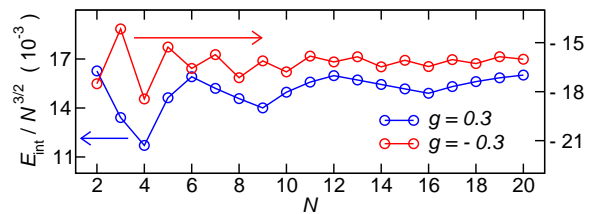}}
\caption{(Colour) 
$E_{\text{int}}/ N^{3/2}$ vs $N$ for $g=\pm 0.3$.
\label{Delta}}
\end{figure}

Figure \ref{Delta} shows the interaction energy $E_{\text{int}}$
for $2\le N\le 20$ for {\it small} $g$ \cite{notesmallg}, where the
non-interacting contribution was subtracted from the total energy,
$E_{\text{int}}=E_0(N)-E_0(N,g=0)$. In order to magnify the fine
structure of $E_{\text{int}}$ in Fig.~\ref{Delta} we replace it
with the scaled quantity $E_{\text{int}}/N^{3/2}$, since in the
Thomas-Fermi approximation $E_{\text{int}}\sim N^{3/2}$. The plot
clearly shows that in the case of repulsive interaction ($g=0.3$)
the energy gain is minimum for closed shells ($N=$ 2, 6, 12, 20).
Conversely, the energy gain is largest for the half-filled open
shells at $N=$ 4, 9, 16. In fact, in those cases the spin is
maximized due to Hund's rule ($S=$ 1, 3/2, 2, respectively). In the
attractive case, already for weak interactions the pattern of
$E_{\text{int}}/N^{3/2}$ shows a marked even-odd alternation: the
energetically favored ground states occur at even $N$ when all atoms
are paired.

\begin{figure}
\centerline{\epsfig{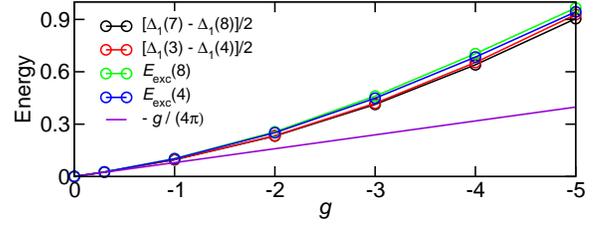}}
\caption{(Colour) 
Alternative estimates of the pairing energy gap $\Delta$ vs
$g$ for $N=$ 4 (3), 8 (7).
\label{DeltaExc}}
\end{figure}

The key quantity for $g<0$ is the pairing energy gap $\Delta$,
measuring the interaction-energy gain (expense) by adding an atom to
the $N$-body system if $N$ is odd (even). We define $\Delta$ as
$\Delta=[\Delta_1(N)-\Delta_1(N+1)]/2$ with $N$ odd. This is plotted
in Fig.~\ref{DeltaExc} vs $g$ for the open-shell cases
$N=3,7$. Remarkably, the black ($N=7$) and red ($N=3$) curves almost
overlap, showing that pairing is a generic feature of the few-body
system.
\begin{figure}
\centerline{\epsfig{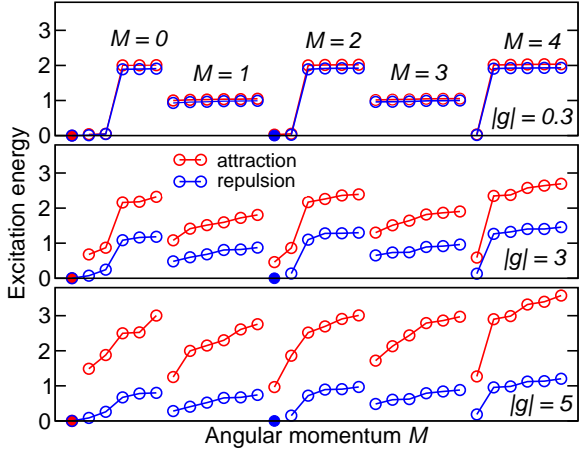}}
\caption{(Colour)  Excitation energies vs $M$ for $N=8$. The top,
middle, bottom panels are obtained for $\left |g\right|=$ 0.3, 3, 5,
respectively. The red (blue) circles label the case of attractive
(repulsive) interaction. Only the six lowest excited states are
shown for each value of $M$. The filled circles mark the ground
states, whose energies are taken as reference. \label{excitation}}
\end{figure}

Complementary information on the pairing energy $\Delta$ is obtained
by the analysis of the excitation spectrum for fixed even $N$. Figure
\ref{excitation} focuses on the low-lying energy levels of $N=8$ as
the magnitude of $g$ increases, going from $\left|g\right|=0.3$ (top
panel) up to $\left|g\right|=5$ (bottom panel). We resolve the six
lowest-energy levels for orbital angular momenta $0\le M \le 4$. For
small values of $g$ ($g = \pm 0.3$, top panel) the non-interacting
low-lying sequence clearly emerges: three $M=0$ states, two $M=2$
states, and one $M=4$ state may be obtained by variously arranging
two atoms with opposite spin in the third open shell of the 2D
harmonic oscillator, while the remaining six atoms fill the two
lowest shells. To have the lowest $M=1$ or $M=3$
excitations one particle (or hole) must be excited into the 4th
(2nd) shell, whereas to have higher-energy states with even $M$, one
particle (hole) must be excited into the 5th (1st) shell or two into
the 4th (2nd) shell. As $\left|g\right|$ increases (center and
bottom panels of Fig.~\ref{excitation}), two fundamentally different
types of excitation spectra appear for positive (blue curves) and
negative (red curves) $g$. For $g<0$, all excited states appear at
some large energy related to the pairing gap $\Delta$: A pair must
be broken to excite the system, no matter the value of $M$. As
$\left|g\right|$ increases, the gap becomes uniform with respect to
$M$, indicating the onset of intershell pairing. The opposite holds
for $g>0$, since excitation energies decrease for increasing $g$,
implying that several excited states become almost degenerate with
respect to the ground state, as in strongly correlated quantum dots
\cite{reimannRMP}.

%
The lowest excitation energy, $E_{\text{exc}}(N)$, may be used as an
alternative definition of $\Delta$ \cite{comment}. Such quantity is
plotted in Fig.~\ref{DeltaExc} for $N=4$ (blue line) and $N=8$
(green line) and may be compared with the previous evaluations of
$\Delta$ through the chemical potential. We see that all the
different estimates basically coincide, independent of $N$. A fair
agreement is found also for cases other than open shells, provided
$\left|g\right|$ is sufficiently large. It is interesting to
compare the curves for $E_{\text{exc}}(N)$ in Fig.~\ref{DeltaExc}
with first-order perturbation theory in $g$
(magenta curve). At small $\left|g\right|$ all estimates of $\Delta$
agree with the predicted value of $-g/(4\pi)$, which is
linear in $g$. Around $g\approx -1$,
however, strong deviations from the perturbation-theory result
arise, due to the non-linear behavior of $\Delta$.

\begin{figure}
\begin{picture}(200,130)
\put(-20,02){\epsfig{file=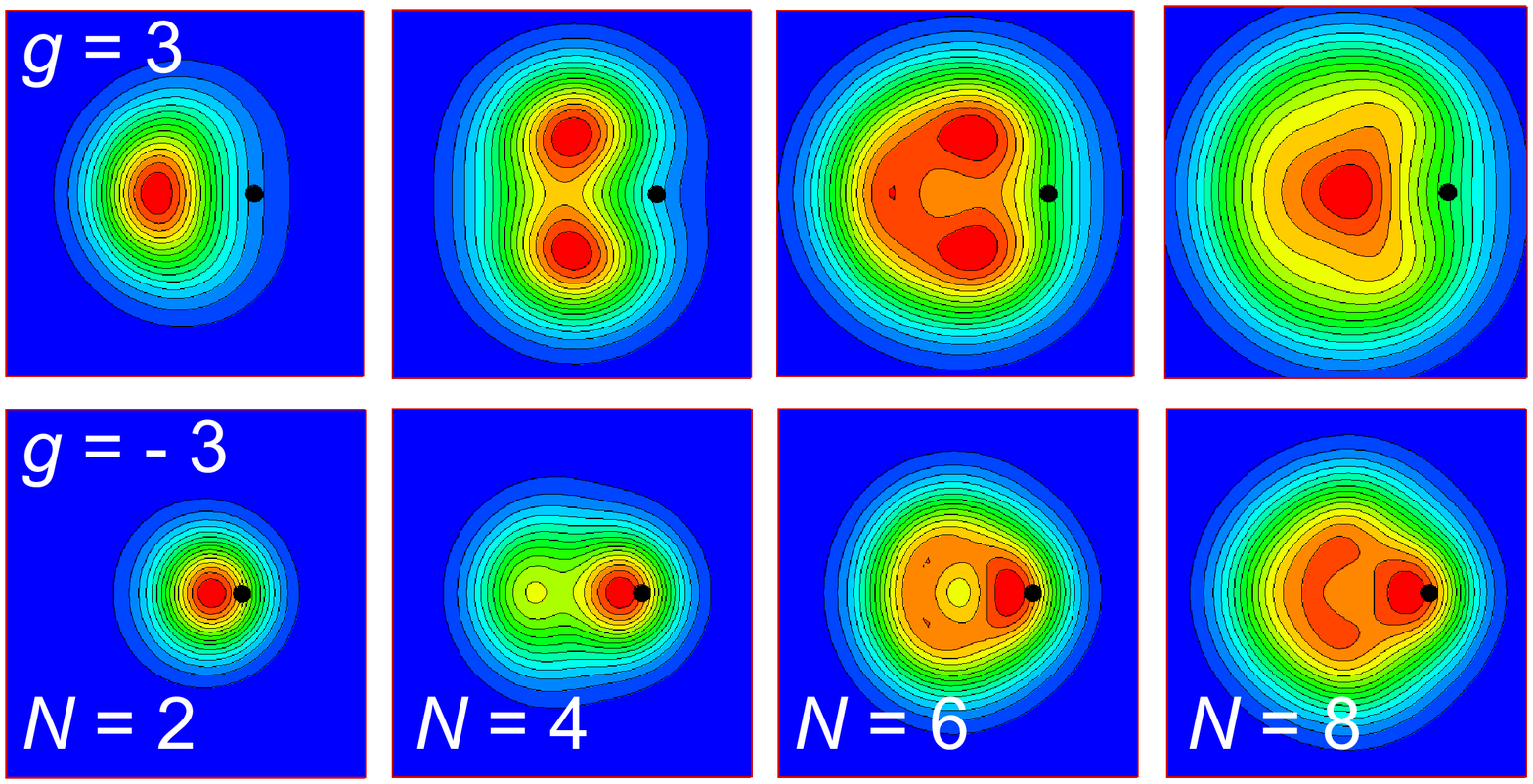,width=2.5in,,angle=0}}
\end{picture}
\caption{(Colour) Plot
of $P_{\uparrow\downarrow}\!(\bm{r},\bm{r}_0)$
in the $xy$ plane. The black dot at $(x,y)=(x_0,0)$ 
locates the spin-$\downarrow$ reference
atom ($x_0$ is the average value of $r$). Top
row: $g=3$. Bottom row: $g=-3$. From left to right column, $N=$ 2,
4, 6, 8, respectively. The squares' size is $5\times 5$, and 15
equally spaced contour levels go from blue (minimum) to red
(maximum). \label{contours}}
\end{figure}

We finally investigate the internal structure of ground
state wave functions, computing the conditional
probability $P_{\sigma\sigma_0}\!(\bm{r},\bm{r}_0)$ of finding
an atom with spin $\sigma= \uparrow,\downarrow$ at
$\bm{r}$ provided another one is fixed at $\bm{r}_0$ with
spin $\sigma_0$:
\begin{equation}
P_{\sigma\sigma_0}\!(\bm{r},\bm{r}_0)=A
\sum_{i\neq j}\left<\delta(\bm{r}-\bm{r}_i)\delta_{\sigma\sigma_i}
\delta(\bm{r}_0-\bm{r}_j)\delta_{\sigma_0\sigma_j}\right>,
\label{eq:P}
\end{equation}
where $A$ is a normalization factor.
The probability $P_{\uparrow\downarrow}\!(\bm{r},\bm{r}_0)$
is plotted in Fig.~\ref{contours} for even $N$, with $\bm{r}_0=(x_0,0)$
labeled by a black circle [for convenience, we scaled the 
maximum height of $P_{\uparrow\downarrow}\!(\bm{r},\bm{r}_0)$
to the same value in each panel].
Figure \ref{contours} only shows
correlations between opposite-spin atoms, since the parallel
spin probability $P_{\sigma_0\sigma_0}\!(\bm{r},\bm{r}_0)$
is almost structureless, except the exchange
hole around the fixed atom (the contact interaction does
not scatter atoms with parallel spin).

Already for $N=2$ (left column of Fig.~\ref{contours}),
the difference between the repulsive (top panel, $g=3$) and
attractive (bottom panel, $g=-3$) case is manifest.
In fact, whereas for $g>0$ the $\uparrow$-atom may be found
in the antipodal position with respect to the fixed $\downarrow$-atom,
for $g<0$ both atoms tend to overlap in space, suggesting the
formation of a bound pair.
The contour plots for $g=-3$ (bottom row) show
a peak of the probability of finding an
$\uparrow$-atom close to the $\downarrow$-atom, and they are
rotated by $2\pi/N$ with respect to
the repulsive case, as displayed in the upper panel.
Intriguingly, whereas the overall spatial distribution for
$P_{\uparrow\downarrow}\!(\bm{r},\bm{r}_0)$ tends to an
isotropic distribution of atoms for $g=3$, for $g=-3$ the
distribution is strongly distorted in space.
This distortion is particularly clear for $N=4$ and $N=6$, with six particles
filling a shell in the non-interacting limit, which should maintain
circular symmetry.
The origin of the deformation seen in $P_{\uparrow\downarrow}$
might be attributed to the
Jahn-Teller effect or the arrangement of bound pairs in the trap,
as discussed by Stecher~{\it et al.}~\cite{stecher2008} for the four-particle
system.

In summary, we studied shell structure and energy gaps for a few fermionic
particles with repulsive as well as attractive contact interactions
in a 2D harmonic trap. For repulsive interactions,
shell structure and Hund's rule lead to a significantly enhanced
fundamental energy gap at closed-shell and mid-shell configurations.
For attractive interactions on the BCS-side of unitarity, however,
a pronounced odd-even staggering was found, where the
\emph{ab-initio} results agree well with the seniority model.
Interaction blockade, as discussed here for 2D
``quantum dots'' with cold atoms, may  experimentally be observed in
atom-transport studies, as recently performed by Cheinet
and coworkers~\cite{cheinet2008} for bosons, and an extension to the 
fermionic case is called for. 

We thank V. Zelevinsky and S. Corni for discussions.
This work was supported by Project FIRB No. 
RBIN04EY74 and RBIN06JB4C,
PRIN No. 2006022932, INFM-CINECA  
Project 2008, the VR, and the SSF.


\begin{thebibliography}{30}

\bibitem{artifiatoms} M. A. Kastner, Physics Today {\bf 46(1),}  24 (1993);
R. C. Ashoori, Nature {\bf 379,} 413 (1996).

\bibitem{reimannRMP} S. M. Reimann and M. Manninen, Rev. Mod. Phys.
{\bf 74,} 1283 (2002).

\bibitem{books} See e.g. L. Pitaevskii and S. Stringari, {\it
Bose-Einstein Condensation} (Clarendon, Oxford, 2003);
C. J. Pethick and H. Smith, {\it Bose-Einstein Condensation in Dilute 
Gases} (Cambridge University Press, Cambridge, 2004).

\bibitem{coldferm} See, e.g., C. A. Regal {\it et al.},
Nature {\bf 424,} 47 (2003); S. Jochim {\it et al.}, Science {\bf
302,} 2101 (2003); M. W. Zwierlein {\it et al.}, Phys. Rev. Lett.
{\bf 91,} 250401 (2003); K. E. Strecker {\it et al.}, Phys. Rev.
Lett. {\bf 91,} 080406 (2003); M. W. Zwierlein {\it et al.}, Nature
{\bf 442,} 54 (2006); Science {\bf 311}, 492 (2006); T. St\"oferle {\it et
  al.}, Phys. Rev. Lett. {\bf 96}, 030401 (2006); J. K. Chin {\it et al.},
Nature {\bf 443}, 961 (2006). 

\bibitem{mottelson} H. Heiselberg and 
B. Mottelson, Phys. Rev. Lett. {\bf 88,}
190401 (2002);
H. Heiselberg, Phys. Rev. A {\bf 68,} 053616 (2003);
G. Bruun and H. Heiselberg, {\it ibid.} {\bf 65,} 053407 (2002).

\bibitem{cheinet2008} S. F\"olling \emph{et al.}, Nature
{\bf 448,} 1029 (2007); P. Cheinet \emph{et al.}, Phys. Rev. Lett.
{\bf 101,} 090404 (2008).

\bibitem{capelle2007} K. Capelle {\it et al.}, Phys. Rev. Lett. {\bf 99,} 
010402 (2007).

\bibitem{seaman2007} B. T. Seaman {\it et al.}, Phys. Rev. A 
{\bf 75,} 023615 (2007).

\bibitem{zinner2008} N. T. Zinner {\it et al.}, arXiv:0803.2861.

\bibitem{RontaniCI} M. Rontani \emph{et al.}, J. Chem. Phys. {\bf 124,} 
124102 (2006).

\bibitem{momentumcutoff} B. D. Esry and C. H. Greene,
Phys. Rev. A {\bf 60,} 1451 (1999); I. Stetcu 
{\it et al.}, {\it ibid.} {\bf 76,} 063613 (2007);
A. Bulgac, J. E. Drut, and P. Magierski, 
Phys. Rev. Lett. {\bf 96,} 090404 (2006);
Y. Alhassid, G. F. Bertsch, and L. Fang,
{\it ibid.} {\bf 100,} 230401 (2008);
I. Stetcu, B. R. Barrett, and U. van Kolck,
Phys. Lett. B {\bf 653,} 358 (2007).  

\bibitem{rontani2008} M. Rontani, S. {\AA}berg,
and S. M. Reimann, arXiv:0810.4305.

\bibitem{Busch} T. Busch {\it et al.},
Found. Phys. {\bf 28,} 549 (1998).

\bibitem{comment2} 
We obtain $a$ = 2.16, 5.25, 358, 8.42 $\times 10^8$, 
5.45 $\times 10^{-10}$, 1.28 $\times 10^{-3}$,
0.0851, 0.198 for
the considered values of $g$ = -5, -3, -1, -0.3, 0.3,
1, 3, 5, respectively. Note that the 2D scattering length is
always positive \cite{Busch}. 

\bibitem{Don} We used the code DONRODRIGO 
for a basis of Slater determinants obtained by filling in the orbitals 
of the lowest 6 shells with $N$ atoms in all possible ways (full CI). 
For $N=8$ the maximum
linear size of the eigenvalue problem was $2.08 \times 10^6$.

\bibitem{tarucha1996} S. Tarucha \emph{et al.}, Phys. Rev. Lett.
{\bf 77,} 3613 (1996).

\bibitem{Jahn-Teller} W. Satu\l{}a,  J. Dobaczewski, and W. Nazarewicz,
Phys. Rev. Lett. {\bf 81,} 3599 (1998); H. H\"akkinen {\it et al.},
{\it ibid.} {\bf 78,} 1034 (1997).

\bibitem{reimann1998} S. M. Reimann {\it et al.}, Phys. Rev. B 
{\bf 58,} 8111 (1998).

\bibitem{volya2001} A. Volya, B. A. Brown, and V. Zelevinsky, Phys. Lett. B
{\bf 509,} 37 (2001); V. Zelevinsky and A. Volya, Phys. At. Nucl. {\bf 66,}
1781 (2003).

\bibitem{notesmallg} Since $g$ is small here we used a space
of 4 shells only.

\bibitem{comment} In BCS and in the seniority model the lowest
excitation energy is $\approx 2\Delta$, while the CI value
is smaller, about $\Delta$.

\bibitem{stecher2008} J. von Stecher, C. H. Greene and D. Blume,
Phys. Rev. A {\bf 77,} 043619 (2008).

\end{thebibliography}
\end{document}